\documentclass{llncs}
 \usepackage[a4paper]{geometry}
\usepackage[utf8]{inputenc}
\usepackage{url}
\usepackage{amsmath,amssymb}
\usepackage{graphicx}

\def\Cplex{{\tt Cplex}}
\def\Gurobi{{\tt Gurobi}}

\def\NN{{\mathbb{N}}}
\def\ZZ{{\mathbb{Z}}}
\def\RR{{\mathbb{R}}}
\def\un{{\rm{1\!\!1}}}
\def\CC{{\mathcal{C}}}

\def\PP{{\mathcal{P}}}

\title{Partial $K$-means with $M$ outliers: Mathematical programs and complexity results}

\author{ Nicolas Dupin\inst{1}[0000-0003-3775-5629] 
\and
 Frank Nielsen\inst{2}[0000-0001-5728-0726]   
}
\institute{Univ Angers, LERIA, SFR MATHSTIC, F-49000 Angers, France\\  \email{nicolas.dupin@univ-angers.fr}
\and
Sony Computer Science Laboratories Inc, Tokyo, Japan\\
 \email{Frank.Nielsen@acm.org}}

\usepackage{xcolor}

\begin{document}

\maketitle
\color{blue}{

\noindent{This paper should be cited as:}
\vskip 0.2cm
\noindent{Dupin, N., Nielsen, F. (2023). Partial K-Means with M Outliers: Mathematical Programs and Complexity Results. In: Dorronsoro, B., Chicano, F., Danoy, G., Talbi, EG. (eds) Optimization and Learning. OLA 2023. Communications in Computer and Information Science, vol 1824. Springer, Cham. \url{https://doi.org/10.1007/978-3-031-34020-8_22} }

}

\color{black}

\begin{abstract}
A well-known bottleneck of Min-Sum-of-Square Clustering (MSSC, the celebrated $k$-means problem) is to tackle the presence of outliers. In this paper, we propose  a Partial clustering variant termed PMSSC which considers a fixed number of outliers to remove. We solve PMSSC by Integer Programming formulations and complexity results extending the ones from MSSC are studied. PMSSC is  NP-hard in Euclidean space  when the dimension or the number of clusters is greater than $2$. Finally, one-dimensional cases are studied: Unweighted PMSSC is polynomial in that case and solved with a dynamic programming algorithm, extending the  optimality property of MSSC with interval clustering. This result holds also for unweighted $k$-medoids with outliers. A weaker optimality property holds for weighted PMSSC, but NP-hardness or not remains
an open question in dimension one.

\vskip 0.15cm
  
\noindent{\textbf{Keywords} : Optimization; Min-Sum-of-Square ; Clustering; $K$-means;
outliers ;   Integer  Programming; Dynamic Programming; Complexity}
\end{abstract}


\section{Introduction}

The $K$-means clustering of $n$ $d$-dimensional points, also called Min Sum of Square Clustering (MSSC) in the operations research community,  is one of  the  most  famous   unsupervised  learning   problem,
and  has been extensively studied in the literature.
MSSC was is known to be NP hard \cite{dasgupta2008hardness} when $d>1$ or $k>1$.
Special cases of MSSC are also  NP-hard in a general Euclidean space:
the problem is still NP-hard when the number of clusters  is $2$~\cite{aloise2009np}, or in  dimension $2$~\cite{mahajan2012planar}.
The case $K=1$ is trivially polynomial. 
The $1$-dimensional (1D) case is polynomially solvable with  a Dynamic Programming (DP) algorithm~\cite{wang2011ckmeans},
with a time complexity in $O(KN^2)$ where $N$ and $K$ are respectively the number of points and  clusters.
This last algorithm was improved in~\cite{gronlund2017fast}, for a
 complexity in $O(KN)$ time using memory  space in $O(N)$.
A famous iterative heuristic to solve MSSC was reported by Lloyd in~\cite{lloyd1982least},
and a local search heuristic is proposed in~\cite{kanungo2002local}.
Many improvements have been made since 
 then: See~\cite{jain2010data} for a review.

 A famous drawback of MSSC clustering is that it is not robust to  noise nor to outliers \cite{jain2010data}.
The $K$-medoid problem, the discrete variant of the $K$-means problem addresses this weakness of MSSC by computing the cluster costs 
by choosing the cluster representative amongs the input points and not by calculating centroids.
Although $K$-medoids is more robust to noise and outliers, it induces more time consuming computations than MSSC
\cite{dupin2019medoids,huang2021comparing}.
In this paper, we define Partial MSSC (PMSSC for short) by considering a fixed number of outliers to remove as in partial versions of facility location problems like
$K$-centers  \cite{dupin2021unified} and $K$-median  \cite{charikar2001algorithms}, and study extensions of exact algorithms of MSSC and report complexity results.
Note that a $K$-means problem with outliers,  studied in  \cite{krishnaswamy2018constant,zhang2021local}, has some similarities with PMMSC, we will precise the difference with PMMSC.
To our knowledge, PMSSC is studied  for the first time in this paper.

The remainder of this paper is structured as follows.
In Section 2, we introduce the notation and formally describe the problem.
In Section 3, Integer Programming formulations are proposed.
In Section 4, we give first complexity results and analyze  optimality properties.
In Section 5, a polynomial  DP algorithm is presented for unweighted MSSC in 1D.
 In Section 6, relations with state of the art  and extension of these result are discussed.
  In Section 7,  our contributions  are summarized,  discussing also future directions of research.
To ease the readability,  the proofs are gathered in an Appendix.

 \section{Problem statement and notation}

Let $E=\{x_1,\dots, x_N\}$ be a set of $N$ distinct elements of $\RR^L$, with $L \in \NN^*$.  
 We note discrete intervals $[\![a,b]\!] = [a,b] \cap \ZZ$, so that we can use the notation of discrete index sets and write $E=\{x_i\}_{i \in [\![1,N]\!]}$.
 We define  $\Pi_K(E)$, as the set of all the possible partitions of $E$ into $K$ subsets:
\begin{equation*}
\Pi_K(E) = \left\{P \subset \PP(E)\: \bigg| \:\forall p,p' \in P, \:\:p \cap p' =  \emptyset \: \mbox{and} \: \bigcup_{p \in P} = E \: \mbox{and} \; \mbox{card}(P)=K \: \right\}
\end{equation*}

MSSC is  special case of K-sum clustering problems.
Defining a  cost function $f$ for each subset of $E$ to measure the dissimilarity,
$K$-sum clustering  are combinatorial optimization problems indexed by $\Pi_K(E)$,
minimizing the sum  of the  measure $f$ for all the $K$ clusters partitioning $E$:
\begin{equation}\label{clusteringGal}
\min_{\pi \in \Pi_K(E)}
\sum_{P \in \pi}  f(P)
\end{equation}

\def\MSSC{\mathrm{MSSC}}
\def\UMSSC{\mathrm{UMSSC}}

\begin{figure}%
\centering
\begin{tabular}{cc}
\includegraphics[width=0.4\columnwidth]{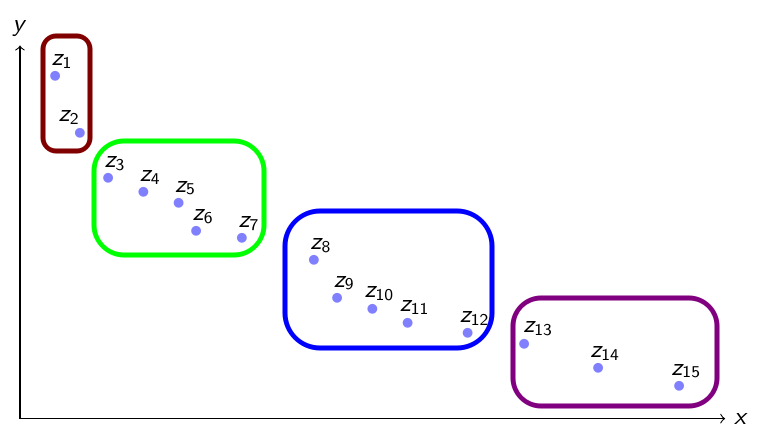}
&
\includegraphics[width=0.4\columnwidth]{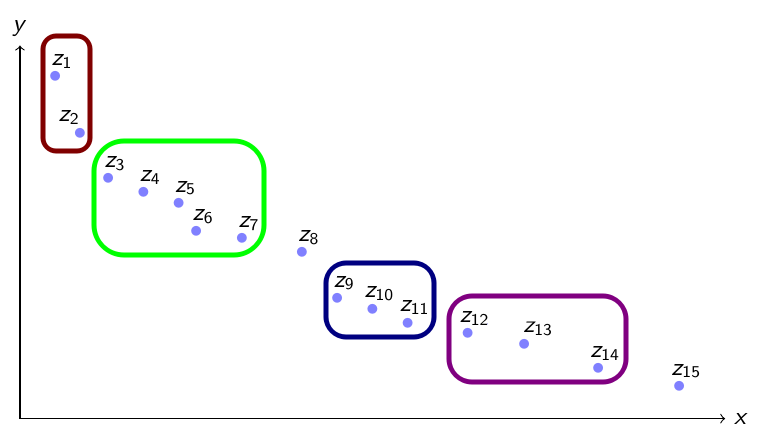} \cr
(a) & (b)\\
\includegraphics[width=0.4\columnwidth]{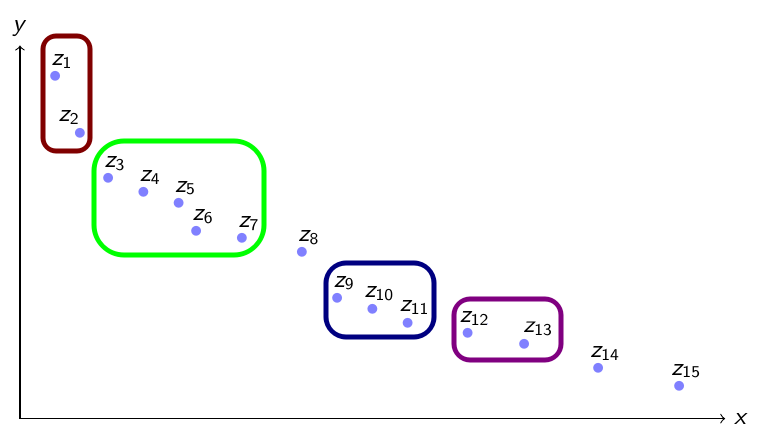}  &
\includegraphics[width=0.4\columnwidth]{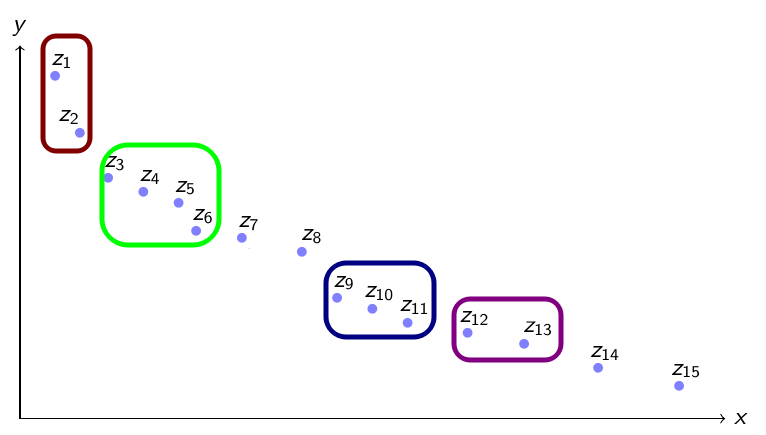} \\
(c) & (d)
\end{tabular}
\caption{MSSC clustering of a Pareto front in 4 clusters: (a) no outliers, (b) 2 outliers, (c) 3 outliers, and (d) 4 outliers. }%
\label{fig:clustering}%
\end{figure}

Unweighted MSSC minimizes the sum for all the $K$ clusters of the average {\em squared distances} from the points of the clusters to the centroid.
Denoting with  $d$ the Euclidean distance in $\RR^L$:
\begin{equation}\label{defMSSC}
\forall P  \subset E, \;\;\;
f_{\UMSSC}(P)=\min_{c \in \RR^L}\sum_{x \in P}  d(x,c)^2 =  \sum_{x \in P}  d\left(x, \frac 1 {|P|} \sum_{y \in P} y\right)^2
\end{equation}
The last equality can be proven using convexity and order one optimality conditions.
In the weighted version, a weight $w_j>0$ is associated to each point $x_j \in E$.
For $x \in E$, $w(x)$ denotes the weight of point $x$.
Weighted version of MSSC considers as dissimilarity function:

\begin{equation}\label{defWeightedMSSC}
\forall P  \subset E, \;\;\;
f_{\MSSC}(P)=\min_{c \in \RR^L}\sum_{x \in P}  w(x) \times d(x,c)^2
\end{equation}

Unweighted cases correspond to $w_j=1$. Analytic computation of weighted centroid holds also with convexity:
\begin{equation}\label{defWeightedMSSC2}
f_{MSSC}(P)= 
\sum_{x \in P}   w(x) \times d\left(x, \frac 1 {\sum_{z \in P} w(z)} \sum_{y \in P} w(y) \times y\right)^2
\end{equation}

We consider a partial clustering extension of MSSC problem,
similarly to the partial p-center and facility location problems \cite{charikar2001algorithms,dupin2021unified}.
A bounded number $M<N$ of the points may be considered outliers and removed in the evaluation.
It is an optimal MSSC computation enumerating each subset $E' \subset E$ removing at most $M$ points, i.e. such that $|E \setminus E'|\leqslant M$. It follows that PMSSC can be written as following combinatorial optimization problem:

\begin{equation}\label{partialMSSC}
\min_{E' \subset E: |E \setminus E'|\leqslant M} \;\min_{\pi \in \Pi_K(E')} \sum_{P \in \pi}  f_{MSSC}(P)
\end{equation}

Figure~\ref{fig:clustering} shows an example of MSSC and PMSSC with $M\in\{2,3,4\}$.

In the "robust $K$-means problem"   studied in \cite{krishnaswamy2018constant,zhang2021local}, also denoted or "$K$-means problem with outliers", "robust" also denotes the partial variant with a defined number of outliers.
It is not the usual meaning of robust optimization in the operations research community.
These papers consider only the unweighted version of the problem, this paper highlights the difficulty of meaningfully formulating such a problem.
The crucial difference with our assumptions is that their partial version concerns a discrete clustering version with a discrete set of possible centroids like $K$-medoids, not a partial version of MSSC where the centroid is continuous. Such problem will be denoted as ``partial $K$-medoids problem", it is  defined with (\ref{partialMSSC})with following $f_{medoids}$ measure instead of $f_{MSSC}$:

\begin{equation}\label{defMedoids}
 f_{medoids}(P)=\min_{c \in P}\sum_{x \in P}  d(x,c)^2
\end{equation}

\section{Mathematical Programming Formulations}\label{sec::mathProg}

Partial MSSC can be formulated with Integer Programming formulations, extending the ones from MSSC \cite{aloise2012improved,peng2007approximating,piccialli2022sos}.

For $n \in [\![1;N]\!]$ and $k \in [\![1;K]\!]$, we use  binary variables
  $z_{n,k} \in \{0,1\}$ defined with  $z_{n,k}=1$ if and only if point $x_n$ is assigned to cluster  $k \in [\![1,K]\!]$.
Using definition (\ref{defWeightedMSSC}),  the weighted centroid of cluster $K$ is defined as a continuous
variable $c_k \in \RR_+^L$. It give rises to a first quadratic formulation:

  \begin{eqnarray}
&  \displaystyle\min_{z_{n,k}, c_k} \displaystyle\sum_{k=1}^K  \sum_{n=1}^N w_n  d(x_n,c_k )^2 z_{n,k} \label{objmathProg}\\
  s.t: &
 \displaystyle \sum_{k=1}^K z_{n,k} \leqslant  1  & \; \; \; \; \forall  n, \label{assignementMathProg0}\\  
 &
 \displaystyle \sum_{n'=1}^N \sum_{k=1}^K z_{n',k} \geqslant  N-M \label{assignementMathProg}
  \end{eqnarray}

Objective function  (\ref{objmathProg}) holds also for (\ref{clusteringGal})
and (\ref{defWeightedMSSC}) with $z_{n,k}$  encoding subsets $P \in \pi$.
If $M=0$, constraint  (\ref{assignementMathProg}) is equivalent to $ \sum_{k=1}^K z_{n',k} =1$
for each index $n'$, point $x_{n'}$ shall be assigned to exactly one cluster.
Constraints  (\ref{assignementMathProg0}) impose that each point is assigned to at least one cluster.
Constraint  (\ref{assignementMathProg}) aggregates that at most $M$ points are unassigned, ie $ \sum_{k=1}^K z_{n',k} =0$ for these  $x_{n'}$,
and the other ones  fulfill $ \sum_{k=1}^K z_{n'',k} =1$ because of
 constraints  (\ref{assignementMathProg0}).
\vskip 0.12cm

 As for unpartial MSSC, last quadratic formulation is not solvable by mathematical programming solvers like \Cplex{} and \Gurobi{} because of non-convexity of the objective function.
 A compact reformulation, as for  unpartial MSSC, allows such straightforward resolution.
 Using additional continuous variables
$s_{n,k} \geqslant 0$ as the squared distance from point $x_n$  to its cluster centroid $c_k$ if $z_{n,k}=1$ and $0$ otherwise.
It induces following convex quadratic formulation with quadratic convex constraints with a big M that can be set to $D=\max_{i,i'} d_{i,i'}^2$:

   \begin{eqnarray}
&  \displaystyle\min_{z_{n,k},s_{n,k}, c_k} \displaystyle\sum_{k=1}^K  \sum_{n=1}^N w_n  s_{n,k}  \label{objmathProg2}\\
  s.t: &
 \displaystyle \sum_{k=1}^K z_{n,k} \leqslant  1  & \; \; \; \; \forall  n, \label{assignementMathProg02}\\  
 &
 \displaystyle \sum_{n'=1}^N \sum_{k=1}^K z_{n',k} \geqslant  N-M \label{assignementMathProg2}\\
  & s_{n,k} \geqslant d(x_n, c_k)^2 - D (1 - z_{n,k}) & \; \; \; \; \forall  n,k, \label{convexMathProg2}
  \end{eqnarray}

%


Previous formulations have a common weakness, it induces symmetric solutions with permutations of clusters,  which makes Branch \& Bound tree search inefficient.
As  in \cite{aloise2012improved} for unpartial MSSC, an extended  reformulation 
can improve this known bottleneck.
 Enumerating each subset of $E$, $p \in \PP=2^E$,  $c_p$ denotes the clustering cost of  $p$ with formula (\ref{defWeightedMSSC2}), and we define a binary variable $z_p \in \{0,1\}$ with $z_p=1$ if and only if subset $p$ is chosen as a cluster.
We define binaries $y_n \in \{0,1\}$ with $y_n=1$ if and only if point $x_n$ is chosen to be counted as outlier and not covered.

  \begin{eqnarray}
\mbox{PMSSC}=
& \displaystyle \min_z \sum_{p \in \PP} c_p z_p  \label{objExt}\\
  s.c: \forall n,\; &  \sum_{p \in \PP} \un_{n \in p} z_p  \geqslant 1 - y_n \label{partialCoverExt} &\\
    &\sum_{n}  y_n \leqslant M  & \label{budgetMext} \\
  &\sum_{p \in \PP}  z_p \leqslant K \label{budgetKext} 
  \end{eqnarray}
Objective function  (\ref{objExt}) is linear in the extended reformulation.
Constraint (\ref{budgetMext}) bounds the maximal budget of uncovered points.
Constraint (\ref{budgetKext}) bounds the maximal number of clusters, having more clusters decreases the objective function.
Constraints (\ref{partialCoverExt}) express that either a point $x_n$ is uncovered
when $y_n=1$ and there is no need  to  select a subset which contains $x_n$, or one subset (at least)
contains  $x_n$. Note that $ \un_{n \in p} z_p $ is one if and only if subset $p$ contains point $x_n$.
These constraints are written  with inequalities, equalities are valid also to have the same optimal solutions.
Inequalities are preferred  for numerical stability with Column Generation (CG) algorithm.

Variables $ z_p$, contrary to variables $y_n$, are  of an exponential size and cannot be enumerated. CG algorithm applies
to generate only a subset of  $z_p$ variables to compute the continuous (LP) relaxation of  (\ref{objExt})-(\ref{budgetKext}) .
We consider the Restricted Master Problem (RMP) for a subset of $z_p$ variables in
  $\PP' \subset \PP$ of the LP relaxation, so that dual variables are defined for each constraint:

\begin{equation}
  \begin{array}{lll}
\mbox{RMP}(\PP')=& \displaystyle \min_{z\geqslant 0} \sum_{p \in \PP'} c_p z_p  \\
  s.c: \forall n,\; & y_n +  \sum_{p \in \PP'} \un_{n \in p} z_p  \geqslant 1& (\pi_n)\\
       &- \sum_{n}  y_n \geqslant - M  & (\lambda)\\
 & - \sum_{p \in \PP'}  z_p \geqslant K & (\sigma)
  \end{array}
\end{equation}

Inequalities imply that dual variables $\sigma, \lambda, \pi_n \geqslant 0$  are signed.
This problem is feasible if $E \in \PP' $ or if a trivial initial solution is given.
Applying strong duality:

\begin{equation}
  \begin{array}{lll}
\mbox{RMP}(\PP')=& \displaystyle \max_{\pi_n,\sigma,\lambda \geqslant 0} - K\sigma - M \lambda + \sum_{n} \pi_n  \\
  s.t: \forall p\in \PP',\; &  - \sigma + \sum_{n} \un_{n \in p} \pi_n  \leqslant c_p & (z_p)\\
  \forall n,\; & \pi_n - \lambda  \leqslant 0 & (y_n)
  \end{array}
\end{equation}

Having only a subset of  $z_p$ variables, RMP is optimal if for the non generated
 $z_p$ variables, we have $ - \sigma + \sum_{n} \un_{n \in p} \pi_n \leqslant c_p$.
Otherwise, a cluster $p$ should be added in the RMP if
$ - \sigma + \sum_{n} \un_{n \in p} \pi_n > c_p$.
It defined CG sub-problems:
\begin{equation}
\mbox{SP} = \min_{p\in \PP} c_p -  \sum_{n} \un_{n \in p} \pi_n
\end{equation}
CG algorithm iterates adding subsets $p$ such that $ c_p -  \sum_{n} \un_{n \in p} \pi_n < - \sigma$.
Once  SP$\geqslant - \sigma$, the RMP is optimal for the full extended formulation.

\vskip 0.23cm
As constraints (\ref{budgetMext}) are always in the RMP, partial clustering  induces the same pricing problem with  \cite{aloise2012improved}.
%
 Primal variables $y_n$ influence numerical values of RMP, and thus the values of dual  variables $\pi_n,\sigma$ that are given to the pricing problem, but not the nature of sub-problems.
Sub-problems SP can be solved with \Cplex{} or \Gurobi{}, using the same reformulation technique as in
 (\ref{objmathProg2})-(\ref{convexMathProg2}). Defining  binaries $z_{n} \in \{0,1\}$ such that  $z_{n}=1$ iff  point $x_n$ is assigned to the current cluster, sub-problem SP is written as:

\begin{equation}
\mbox{SP} = \min_{p\in \PP} c_p -  \sum_{n}  \pi_n z_n
\end{equation}
Considering continuous variables $c \in \RR^d$ for the  centroid of the optimal cluster,
and  $s_{n} \geqslant 0$,  the squared  distance from point $x_n$  to centroid $c$ if $z_{n}=1$ and $0$ otherwise. It gives rise to the following
convex quadratic formulation:

\begin{equation}
  \begin{array}{ll}
\mbox{SP} = & \displaystyle\min_{z_{n},s_{n}, c_d}  \displaystyle  \sum_{n=1}^N  s_{n} -  \sum_{n}  \pi_n z_n  \\
  s.t:
  \forall  n, \;\;
  & s_{n} \geqslant d(x_n, c)^2 - D (1 - z_{n})
  \end{array}
\end{equation}

CG algorithm can thus be implemented using \Cplex{} or \Gurobi{}
for LP computations of RMP and for computations of SP.
This gives a lower bound of the integer optimum.
Integer optimality can be obtained using  Branch \& Price. 

\section{First complexity results, interval clustering properties}

PMSSC polynomially reduces to  MSSC: if any instance of PMSSC (or  a subset of instances) is polynomially solvable, this is the case for any corresponding instance of MSSC considering the same points and a value $M=0$ and the same  algorithm.
Hence, NP-hardness results from  \cite{aloise2009np,dasgupta2008hardness,mahajan2012planar} holds for PMSSC:

\begin{theorem}\label{thm:NP-hardness}
Following NP-hardness results holds for PMSSC:
\begin{itemize}
 \item[$\bullet$] PMSSC is  NP-hard for general instances.
 \item[$\bullet$] PMSSC is   NP-hard in a general Euclidean space.
 \item[$\bullet$] PMSSC is NP-hard for instances with a fixed value of $K\geqslant 2$.
\item[$\bullet$] PMSSC is NP-hard for instances with  a fixed value of  $L\geqslant 2$.
 \end{itemize}
\end{theorem}

After Theorem \ref{thm:NP-hardness}, it remains to study cases $K=1$ and $L=1$, where MSSC is polynomial.
 In the remainder of this paper, we suppose that $L=1$, ie we consider the 1D case.
 Without loss of generality in 1D, we consider $d(x,y) = |x-y|$.
We suppose that $E= \{x_1 <\dots< x_N \}$, a sorting procedure running  in $O(N \log N)$ time may be applied.
A key element for the polynomial complexity of MSSC is the interval clustering property \cite{nielsen2014optimal}:

\begin{lemma}\label{lem:intervalClusteringMSSC}
Having $L=1$ and $M=0$, each global minimum of MSSC is only composed of clusters  $\CC_{i,i'} =
\{x_j\}_{j \in [\![i,i']\!]}= \{x \in E \: | \: \exists j \in [\![i,i']\!],\: x = x_j \} $.
\end{lemma}

The question is here to extend this property for PMSSC.
Considering an optimal solution of PMSSC the restriction to no-outliers points is an optimal solution of PMSSC and an interval clustering property holds:

\begin{proposition}\label{prop:intervalClusteringPMSSC}
Having $L=1$ and an optimal solution  of  PMSSC induce an
optimal solution of MSSC removing the outliers. In this subset of points, the optimality property of interval clustering holds.
\end{proposition}

Proposition \ref{prop:intervalClusteringPMSSC} is weaker than Lemma \ref{lem:intervalClusteringMSSC},
selected points are not necessarily an interval clustering with the  indexes of $E$.
This stronger property is false in general for weighted PMSSC, one can have optimal solutions with outliers to remove
inside the natural interval cluster as in the following example with $M=1$, $L=1$  and $K=2$:\label{cex}

$\bullet$ $x_1 = 1$, $w_1 = 10$

$\bullet$ $x_2=2$, $w_2=1000$

$\bullet$ $x_3=3$, $w_2=1$

$\bullet$ $x_4=100$,  $w_4=100$

$\bullet$ $x_5=101$,  $w_5=1$

Optimal PMSSC consider $x_2$ as outlier,  $\{x_1;x_3\}$ and $\{x_4;x_5\}$ as the two clusters.
For $K=1$, changing the example with $x_4=3.001$ and $x_5=3.002$, gives also a counter example with $K=1$ with
$\{x_1;x_3;x_4;x_5\}$ being the unique optimal solution.
These counter-examples 
use a significant difference in  the weights.
In the unweighted PMSSC, interval property holds as in  Lemma \ref{lem:intervalClusteringMSSC}, with outliers (or holes) between the original interval clusters:

%

\begin{proposition}\label{prop:intervalClusteringUnweightedPMSSC}
 Having $L=1$, each global minimum of unweighted PMSSC is only composed of clusters  $\CC_{i,i'} =
\{x_j\}_{j \in [\![i,i']\!]}$. 
In other words, the $K$ clusters may be indexed  $\CC_{i_1,j_1}, \dots, \CC_{i_K,j_K}$ with
$1 \leqslant i_1 \leqslant j_1 < i_2 \leqslant j_2 <  \dots <  i_K \leqslant j_K\leqslant N$
and $  \sum_{k=1}^{K}  (j_k-i_k)  \geqslant N -  M - K$.
\end{proposition}

%


\vskip 0.3cm

As in \cite{dupin2019medoids}, the efficient computation of cluster cost is a crucial element to compute
the polynomial complexity. Cluster costs can be computed from scratch, leading to polynomial algorithm.
Efficient cost computations use inductive relations for amortized computations in $O(1)$ time, extending the relations  in \cite{wang2011ckmeans}.
We define for $i,i'$ such that $1 \leqslant i \leqslant i' \leqslant  N$:
\begin{itemize}
 \item[$\bullet$] $\displaystyle b_{i,i'} =  \sum_{k=i}^{i'}  \frac {w_k} {\sum_{l=i}^{i'}  w_{l}}    x_k$ the weighted centroid of  $\CC_{i,i'}$.
\item[$\bullet$] $ c_{i,i'}= \sum_{j=i}^{i'} w_j d(x_j , b_{i,i'})^2$ the weighted cost of cluster $\CC_{i,i'}$.
\item[$\bullet$] $v_{i,i'}= \sum_{j=i}^{i'} w_j$
\end{itemize}

\begin{proposition}\label{prop:inductionCostCluster}
Following induction relations holds to compute efficiently $b_{i,i'},v_{i,i'}$ with amortized $O(1)$ computations:
\begin{eqnarray}
\label{inducCumulatedWeight1}
 v_{i,i'+1}  = & w_{i'+1} +  v_{i,i'} &, \;\;\; \forall 1 \leqslant i \leqslant i' < N\\
\label{inducCumulatedWeight2}
 v_{i-1,i'}  = & w_{i-1} +  v_{i,i'}&, \;\;\; \forall 1 <i \leqslant i'  \leqslant  N\\
\label{inducCentroid1W}
 b_{i,i'+1}   = & \dfrac {w_{i'+1} x_{i'+1} + b_{i,i'}  v_{i,i'}} {v_{i,i'+1}} &, \;\;\; \forall 1 \leqslant i \leqslant i' < N\\
\label{inducCentroid2W}
 b_{i-1,i'}  = & \dfrac {w_{i-1} x_{i-1} + b_{i,i'} v_{i,i'}} { v_{i-1,i'}} &, \;\;\; \forall 1 <i \leqslant i'  \leqslant  N
 \end{eqnarray}
Cluster costs are then computable with amortized $O(1)$ computations:
 \begin{eqnarray}
\label{inducCost1W}
 c_{i,i'+1}  = & c_{i,i'} + w_{i'+1} (x_{i'+1} - b_{i,i'})^2 +  v_{i,i'} (b_{i,i'+1} - b_{i,i'})^2\\
\label{inducCost2W}
 c_{i-1,i'}  = & c_{i,i'} + w_{i-1} (x_{i-1} - b_{i,i'})^2 +  v_{i,i'} (b_{i-1,i'} - b_{i,i'})^2
 \end{eqnarray}

Trivial relations $v_{i,i} = w_i$, $b_{i,i} = x_i$ and $c_{i,i}=0$ are terminal cases. 
\end{proposition}

Proposition  \ref{prop:inductionCostCluster} allows to prove 
Propositions
\ref{prop:computationCost1} and \ref{prop:computationCost2} to compute efficiently cluster costs.
Proposition  \ref{prop:inductionCostCluster} is also a key element to have first complexity results with $K=1$ and  $M \leqslant 1$ in Propositions \ref{prop:complexityUnweightedK=1}, \ref{prop:complexityWeightedK=1M=1}.

\begin{proposition}\label{prop:computationCost1}
Cluster costs $c_{1,i}$ for all $i \in [\![1;N]\!]$ can be computed in $O(N)$ time using $O(N)$ memory space.
\end{proposition}

\begin{proposition}\label{prop:computationCost2}
For each  $j \in [\![1;N]\!]$
cluster costs $c_{i,j}$ for all $i \in [\![1;j]\!]$ can be computed in $O(j)$ time using $O(j)$ memory space.
\end{proposition}

\begin{proposition}\label{prop:complexityUnweightedK=1}
 Having $L=1$ and $K=1$, unweighted PMSSC is solvable in $O(N)$ time using $O(1)$ additional memory space.
\end{proposition}

\begin{proposition}\label{prop:complexityWeightedK=1M=1}
 Having $L=1$, $M=1$ and $K=1$, weighted PMSSC is solvable in $O(N)$ time using $O(N)$ memory space.
\end{proposition}

\section{DP polynomial algorithm for 1D unweighted PMSSC}\label{sec::DP1D}

Proposition \ref{prop:intervalClusteringUnweightedPMSSC} allows to design a DP algorithm for unweighted PMSSC, extending the one from \cite{wang2011ckmeans}.
 We define $O_{i,k,m}$ as the optimal cost of unweighted PMSSC with $k$ clusters  among points $[\![1,i]\!]$ with a budget of $m$ outliers for all $i \in [\![1,N]\!]$, $k \in [\![1,K]\!]$ and  $m \in [\![0,M]\!]$.
Proposition \ref{bellman} sets induction relations allowing to compute all the
$O_{i,k,m}$, and in particular $O_{N,K,M}$:

\begin{proposition}[Bellman equations]
\label{bellman}
 Defining $O_{i,k,m}$ as the optimal cost of unweighted MSSC  among points $[\![1,i]\!]$ for all $i \in [\![1,N]\!]$ , $k \in [\![1,K]\!]$ and $m \in [\![0,M]\!]$, we have the following induction relations
\begin{equation}\label{inducInit0}
\forall i \in [\![1,N]\!], \:\:\: O_{i,1,0} = c_{1,i}
\end{equation}
\begin{equation}\label{inducInit2}
\forall  m \in [\![1,M]\!], \: \forall k \in [\![1,K]\!],  \: \forall i \in [\![1,m+k]\!], \:\:\: O_{i,k,m} = 0
\end{equation}
\begin{equation}\label{inducInit3}
\forall m \in [\![1,M]\!], \: \forall  i \in [\![m+2,N]\!],  \:\:\: O_{i,1,m} = \min\left(O_{i-1,1,m-1},c_{1+m,i})\right)
\end{equation}
\begin{equation}\label{inducForm}
 \forall k \in [\![2,K]\!], \:\forall i \in [\![k+1,N]\!], \:\:\: O_{i,k,0}= \min_{j \in [\![k,i]\!]}  \left(O_{j-1,k-1,0} +c_{j,i}\right)
\end{equation}

$\forall  m \in [\![1,M]\!], \: \forall k \in [\![2,K]\!], \: \forall i \in [\![k+m+1,N]\!],$\\
\begin{equation}\label{inducForm2}
O_{i,k,m}= \min \left( O_{i-1,k,m-1}   ,  \min_{j \in [\![k+m,i]\!]}  \left(O_{j-1,k-1,m}+ c_{j,i}  \right)  \right)
 \end{equation}
\end{proposition}

Using Proposition \ref{bellman},  a recursive and memoized   DP algorithm can be implemented
to solve unweighted PMSSC in 1D.
Algorithm 1 presents a sequential implementation, iterating with  index $i$ increasing.
The complexity analysis of Algorithm 1 induces Theorem \ref{Thm1}, unweighted PMSSC is polynomial in 1D.

\begin{figure}[h]
\begin{tabular}{ l }
\hline
\textbf{Algorithm 1:}  DP algorithm for unweighted PMSSC in 1D\\
\hline

\verb!  ! sort $E$ in the increasing order \\
\verb!  ! initialize 
$O_{i,k,m}:=0$  for all $m\in [\![0;M]\!],k\in [\![1;K-1]\!], i\in [\![k;N-K +k]\!]$\\
\verb!  ! compute $c_{1,i}$ for all $i \in [\![1;N-K +1]\!]$ and store in  $O_{i,1,0} :=  c_{1,i}$\\
\verb!  ! \textbf{for} $i:=2$ to $N$\\
\verb!    ! compute and store  $c_{i',i}$ for all $i' \in [\![1;i]\!]$ \\ 
\verb!    ! compute  $O_{i,k,0}: = \min_{j \in [\![k,i]\!]}  \left(O_{j-1,k-1,0} +c_{j,i}\right)$ {for all } $k\in [\![2 ; \min(K,i)]\!]$\\
\verb!    ! \textbf{for} $m=1$ to $\min(M,i-2)$\\
\verb!       ! compute $O_{i,1,m} := \min\left(O_{i-1,1,m-1},c_{1+m,i}\right)$\\
\verb!       ! \textbf{for} $k=2$ to $\min(K,i-m)$\\
\verb!          ! compute  $O_{i,k,m}:= \min \left( O_{i-1,k,m-1}   ,  \min_{j \in [\![k+m,i]\!]}  \left(O_{j-1,k-1,m} +c_{j,i}\right)   \right)$\\
\verb!       ! \textbf{end for} \\
\verb!    ! \textbf{end for} \\
\verb!    ! delete the stored  $c_{i',i}$ for all $i' \in [\![1;i]\!]$\\ 
\verb!  ! \textbf{end for} \\
\verb!  ! initialize $\PP=\emptyset$, $\underline{i}=\overline{i}=N$, $m=M$\\
\verb!  ! \textbf{for} $k=K$ to $1$ with increment $k \leftarrow k-1$\\
\verb!    ! compute $\overline{i}:= \min\{i \in [\![\underline{i}-m;\underline{i}]\!] | O_{\underline{i},k,m} := O_{\underline{i}-i,k,m-i+\underline{i}}  \}$\\
\verb!    ! $m:= m-\overline{i}+\underline{i} $\\
\verb!    ! compute and store  $c_{i',\overline{i}}$ for all $i' \in [\![1;\overline{i}]\!]$\\ 
\verb!    ! find $\underline{i}\in [\![1,\overline{i}]\!]$ such that $\underline{i}:=  \mbox{arg}\min_{j \in [\![k+m,i]\!]}  \left(O_{j-1,k-1,m} +c_{j,\overline{i}}\right) $\\
\verb!    ! add $[x_{\underline{i}},x_{\overline{i}}]$ in $\PP$\\
\verb!    ! delete the stored  $c_{i',\overline{i}}$ for all $i' \in [\![1;\overline{i}]\!]$\\
\verb!  ! \textbf{end for} \\
\textbf{return}  $O_{N,K,M}$ the optimal cost  and the  selected clusters $\PP$ \\
\hline
\end{tabular}
\end{figure}

\begin{theorem}\label{Thm1}
Unweighted PMSSC is polynomially solvable in 1D,
Algorithm 1 runs  in  $O(KN^2(1+M))$ time and use  $O(KN(1+M))$ memory space
to solve unweighted 1D instances of  PMSSC.
\end{theorem}


\section{Discussions}

\subsection{Relations with state of the art results for 1D instances}

Considering the 1D standard MSSC with $M=0$, the complexity of Algorithm 1 is identical with the one from  \cite{wang2011ckmeans},
it is even the same DP algorithm in this sub-case  written using weights.
The partial clustering extension implied using a $M+1$ time bigger DP matrix, multiplying by $M$ the time and space complexities.
This had the same implication in the complexity for p-center problems \cite{dupin2019planar,dupin2021unified}.
Seeing Algorithm 1 as an extension of  \cite{wang2011ckmeans}, it is a perspective to analyze if some improvement techniques for time and space complexity  are valid for PMSSC.

As in  \cite{dupin2021unified}, a question is to define a proper value of $M$ in PMSSC.
Algorithm 1 can give all the optimal $O_{N,K,m}$ for $m \leqslant M$, for a good trade-off decision.
From a statistical standpoint, a given percentage of outliers may be considered.
If we consider that $1 \%$ (resp $5 \%$) of the original points may be outliers, it induces   $M= 0,01 \times N$ (resp $M= 0,05 \times N$). In these cases, we have $M=O(N)$ and the asymptotic  complexity of Algorithm 1 is in
$O(KN^3)$ time and using  $O(KN^2)$ memory space. If this remains polynomial,  this cubic complexity becomes a bottleneck for large vales of $N$ in practice.

In  \cite{dupin2021unified}, partial min-sum-k radii has exactly the same complexity when $\alpha = 2$,
which is quite comparable to PMSSC but considering only the extreme points of clusters with squared distances.
PMSSC is more precise with a weighted sum than considering only the extreme points,
having equal complexities induce to prefer partial MSSC for the application discussed in  \cite{dupin2021unified}.
A reason is that the $O(N^2)$ time computations of cluster costs are amortized in the DP algorithm.
Partial min-sum-k radii has remaining advantages over PMSSC: cases $\alpha=1$ are solvable in $O(N \log N)$ time
and the extension is more general than 1D instances and also valid in a planar Pareto Front  (2D PF).
It is a perspective to study PMSSC for 2D PFs, Figure~\ref{fig:clustering} shows in that case that it makes sense to consider an extended interval optimality as in  \cite{dupin2019medoids,dupin2021unified}.

%
%
%

\subsection{Definition of weighted PMSSC}

Counter-example of Proposition \ref{prop:intervalClusteringPMSSC} page \pageref{cex} shows that considering both (diverse) weights and partial clustering as defined in  (\ref{partialMSSC}) may not remove outliers, which was the motivating property. This has  algorithmic consequences,  Algorithm 1 and the optimality property are specific to unweighted cases.
One can wonder the sense of weighted and partial clustering after such counter-example,
and if  alternative definitions exist.

Weighted MSSC can be implied by an aggregation of very similar points, the weight to the aggregated point being the number of original points aggregated in this new one. This can speed-up heuristics for MSSC algorithms.
In this case, one should consider a budget of outliers $M$, which is weighted also by the points.
Let $m_n$ the contribution of a point $x_n$ in the budget of  outliers.
(\ref{partialMSSCnew?}) would be the definition of partial MSSC with budget instead of (\ref{partialMSSC}):

\begin{equation}\label{partialMSSCbudgetDef}
X =   \left \{E' \subset E: \sum_{ x_n \in E \setminus E'} m_n x_n|\leqslant M \right\}
\end{equation}
\begin{equation}\label{partialMSSCnew?}
\min_{x \in X} \;\min_{\pi \in \Pi_K(x)} \sum_{P \in \pi}  f(P)
\end{equation}

 (\ref{partialMSSC}) is a special case of (\ref{partialMSSCnew?})  considering $m_n=1$ for each $n \in [\![1;N]\!]$.
Note that this extension is compatible with the developments of Section \ref{sec::mathProg}, replacing respectively constraints (\ref{assignementMathProg}) and
(\ref{budgetMext})
by linear constraints (\ref{assignementMathProgVar}) and
(\ref{budgetMextVar}).
These new constraints are still linear, there are also compatible with the convex quadratic program and the CG algorithm for the extended formulation:
\begin{eqnarray}
 \sum_{n'=1}^N \left(1 - m_{n'}  \sum_{k=1}^K z_{n',k} \right) \geqslant &  M  \label{assignementMathProgVar}\\
\sum_{n=1}^N m_n  y_n \leqslant & M \label{budgetMextVar}
 \end{eqnarray}

For the DP algorithm of section \ref{sec::DP1D}, we have to suppose  $m_n \in \NN$. Note that it is the case with aggregation of points,
fractional or decimal $m_n$ are equivalent to this hypothesis, it is not restrictive.
Bellman equations can be adapted in that goal:
(\ref{inducInit2}), (\ref{inducInit3}) and  (\ref{inducForm2}) should be replaced by:
\begin{equation}\label{inducInit2var}
\forall  m \in [\![1,M]\!], \: \forall k \in [\![1,K]\!],  \: \forall i,  \:\:\:  \sum_{j=1}^i m_i \leqslant m \Longrightarrow  O_{i,k,m} = 0
\end{equation}
\begin{equation}\label{inducInit3var1}
\forall m \in [\![1,M]\!], \: \forall  i,  \:\:\:  m_i > m \Longrightarrow  O_{i,1,m} = c_{\alpha_m,i}
\end{equation}
\begin{equation}\label{inducInit3var2}
\forall m \in [\![1,M]\!], \: \forall  i,  \:\:\:  m_i \leqslant m \Longrightarrow  O_{i,1,m} = \min\left(O_{i-1,1,m-m_i},c_{\alpha_m,i}\right)
\end{equation}

where $\alpha_m$ is 
the minimal index such that $ \sum_{j=1}^{\alpha_m} m_j > m$.
\begin{eqnarray}\label{inducForm2Var}
&&m_i \leqslant m \Longrightarrow    O_{i,k,m}= \min \left( O_{i-1,k,m-m_i}   ,  \min_{j \in [\![1,i]\!]}  \left(O_{j-1,k-1,m}+ c_{j,i}  \right)  \right)\\
&&m_i > m \Longrightarrow    O_{i,k,m}=   \min_{j \in [\![1,i]\!]}  \left(O_{j-1,k-1,m}+ c_{j,i}  \right)
 \end{eqnarray}

This does not change the complexity of the DP algorithm.
However,  we do not have necessarily  the property $M<N$ anymore.
In this case, DP algorithm in 1D is pseudo-polynomial.

\subsection{From exact 1D DP to DP heuristics?}

If hypotheses $L=1$ and unweighted PMSSC are restrictive, Algorithm 1 can be used in a DP heuristic with more general hypotheses.
In dimensions $L \geqslant 2$, a projection like Johnson-Lindenstrauss or linear regression in 1D, as in \cite{huang2021comparing},
 reduces heuristically the original problem, solving it with  Algorithm 1 provides  a heuristic clustering solution by re-computing the cost in the original space.
This may be efficient for 2D PFs, extending results from  \cite{huang2021comparing}.

Algorithm 1 can be used with weights. For the cost computations, Propositions \ref{prop:computationCost1} and \ref{prop:computationCost2} make no difference in complexity. 
Algorithm 1 is not necessarily optimal in 1D in the unweighted case,
it gives  the best solution with interval clustering, and no outliers inside clusters. It is a primal heuristic, it furnishes feasible solutions. One can refine this heuristic considering also the possibility of having at most one outlier inside a cluster. Let $c_{i,i'}^{(0)}$ be the cost of cluster $x_i,\dots,x_{i'}$ as previously and also $c_{i,i'}^{(1)}$ the best cost of clustering $x_i,\dots,x_{i'}$ with one outlier inside  that can be computed  as in Proposition \ref{prop:complexityWeightedK=1M=1}.
The only adaptation of Bellman equations that would be required is to replace (\ref{inducInit3}, (\ref{inducForm2}) by:
\begin{equation}\label{inducInit3new}
\forall m \in [\![1,M]\!], \: \forall  i \in [\![m+2,N]\!],  \:\:\: O_{i,1,m} = \min\left(O_{i-1,1,m-1},c_{1+m,i}^{(0)},c_{1+m,i}^{(1)})\right)
\end{equation}
$\forall  m \in [\![1,M]\!], \: \forall k \in [\![2,K]\!], \: \forall i \in [\![k+m+1,N]\!],$\\
\begin{equation}\label{inducForm2new}
O_{i,k,m}= \min \left( O_{i-1,k,m-1}   ,  \min_{j \in [\![k+m,i]\!], l \in \{0,1\}}  \left(O_{j-1,k-1,m-l}+ c_{j,i}^{(l)}  \right) \right)
 \end{equation}

 Note that if case $L=1$ and $K=1$ is proven polynomial, one may compute in polynomial time $c_{j,i}^{(m)}$ values of optimal clustering with $m$ outliers with points indexed in $[\![j,i]\!]$ and solve weighted PMSSC in 1D with similar Bellman equations. This is still an open question after this study.

  \subsection{Extension to partial $K$-medoids}

  In this section, we consider the  partial  $K$-medoids problem with $M$ outliers defined by (\ref{partialMSSC}) and (\ref{defMedoids}), as in  \cite{krishnaswamy2018constant,zhang2021local}.
  To our knowledge, the 1D sub-case was not studied, a minor adaptation of our results and proofs allows to prove this sub-case is polynomially solvable.
Indeed,  Lemma \ref{lem:intervalClusteringMSSC} holds with  $K$-medoids as proven in   \cite{dupin2019medoids}.
 Propositions \ref{prop:computationCost1} and \ref{prop:computationCost2}  have their equivalent in  \cite{dupin2018clustering}, complexity of such operations being in $O(N^2)$ time instead of  $O(N)$ for MSSC.
Propositions \ref{prop:intervalClusteringPMSSC} and  \ref{prop:intervalClusteringUnweightedPMSSC}  still hold with the same proof for $K$-medoids.
 Proposition \ref{bellman} and Algorithm 1 are still valid with the same proofs, the only difference being the different computation of cluster costs.
In Theorem \ref{Thm1} this only changes the time complexity: computing the cluster costs $c_{i,i'}$ is  in $O(N^3)$ time instead of $O(N^2)$, it is not bounded by  the  $O(KN^2(1+M))$ time  to compute the DP matrix.
This results in the theorem:

 \begin{theorem}\label{Thm3}
Unweighted partial  $K$-medoids problem with $M$ outliers is polynomially solvable in 1D,
1D instances are solvable  in  $O(N^3 + KN^2(1+M))$ time and using  $O(KN(1+M))$ memory space.
\end{theorem}

\section{Conclusions and perspectives}

To handle the problem of MSSC clusters with outliers, we introduced in this paper partial clustering variants for unweighted and weighted MSSC.
This problem differs from the  "robust $K$-means problem" (also noted "$K$-means problem with outliers"), which consider discrete and enumerated centroids unlike MSSC.
 Optimal solution of weighted PMSSC may differ from intuition of outliers: We discuss about this problem  and present another similar variant.
 For these extensions of MSSC, mathematical programming formulations for solving exactly MSSC can be generalized.
Solvers like \Gurobi{} or \Cplex{} can be used for a compact and an extended reformulation of the problem.
 NP-hardness results of these generalized MSSC problems holds.  
Unweighted PMSSC is polynomial in 1D and solved with a dynamic programming algorithm which relies on the  optimality property of interval clustering.
With small adaptations, "$K$-means problem with outliers" defined as the
unweighted partial  $K$-medoids problem with $M$ outliers is also  polynomial in 1D and solved with a similar  algorithm.
We show that a weaker optimality property holds for weighted PMSSC. 
The relations with similar  state-of-the-art results and adaptation of the DP algorithm to DP heuristics  are also discussed.

This work opens perspectives to solve this new PMSSC problem.
 The NP-hardness complexity of weighted  PMSSC for 1D instances is still an  open question. 
Another perspective is to extend 1D polynomial DP algorithms for PMSSC for 2D PFs, as in \cite{dupin2019medoids,dupin2021unified}.
Approximation results may be studied for  PMSSC also, trying to generalize  results from \cite{krishnaswamy2018constant,zhang2021local}.
Using only quick and efficient heuristics without any guarantee  would be sufficient for an
application to evolutionary algorithms to detect isolated points in PFs, as in \cite{dupin2021unified}.
Adapting local search heuristics for PMSSC  is also another perspective \cite{huang2021comparing}.
If $K$-medoids variants with or without outliers are used to induce more robust clustering to noise and outliers, the use of PMSSC is promising to retain this property without having slower calculations of cluster costs with $ K$-medoids.
Finally, using PMSSC as a heuristic for $K$-medoids is also a promising venue for future research.

\bibliographystyle{abbrv}
\bibliography{biblio.bib}

\section*{Appendix: proofs of intermediate results}

\noindent{\textbf{Proof of Lemma \ref{lem:intervalClusteringMSSC}}:}
We prove the result by induction on $K \in \NN$.
For $K=1$, the optimal cluster is $E = \{x_j\}_{j \in [\![1,N]\!]}$.
Note that $N \leqslant K$ is also a trivial case, we
suppose $1<K< N$ and the Induction Hypothesis (IH) that   Lemma \ref{lem:intervalClusteringMSSC}  is true for $K-1$.
Let an optimal clustering partition, denoted with clusters $\CC_1, \dots, \CC_K$ and centroids
$c_1 < \dots <c_K$, where $c_i$ is the centroid of cluster $\CC_i$. Strict inequalities are a consequence of Lemma \ref{lem:differentCentroidsOptim}.
Necessarily,  $x_N \geqslant c_K$ and $x_N  \in \CC_K$ because $x_N$ is assigned to the closest centroid.
Let $A = \{ i \in [\![1,N]\!] \:| \: \forall k \in  [\![i,N]\!], x_k \in \CC_K \}$ and
let $j = \min A$. 
If $j=1$,  $E=\CC_K= \{x_j\}_{j \in [\![1,N]\!]}$, it is in contradiction with $K>1$.
 $j-1 \in A$ is  in a contradiction with $j = \min A$.
Hence, we suppose $j>1$ and $j-1 \notin A$ .
Necessarily $x_{j-1} \in \CC_{K-1}$, $c_{K-1}$ is the closest centroid among $c_{1}, \dots, c_{K-1}$.
For each $l \in [\![1,j-2]\!]$, $x_l$ is strictly closer from centroid $c_{K-1}$ than from  centroid $c_{K}$, then  $x_{l} \notin \CC_{K}$ and $A = [\![j,N]\!]$.
 On one hand, it implies  that $\CC_K=\{x_l\}_{l \in [\![j,N]\!]}$.
 On the other hand, the other clusters are optimal  for $E' = E \setminus \CC$ with weighted $(K-1)$-means clustering.
 Applying 
 IH 
 proves that the optimal clusters
 are of the shape $\CC_{i,i'} = \{x_j\}_{j \in [\![i,i']\!]}$.
 \hfill$\square$
\vskip 0.2cm
\begin{lemma} \label{lem:differentCentroidsOptim}
We suppose $L=1$ and $K < N$. Each global optimal solution of weighted MSSC indexed with clusters $\CC_1, \dots, \CC_K$ and centroids such that
$c_1 \leqslant \dots \leqslant c_K$, where $c_i$ is the centroid of cluster $\CC_i$, fulfills necessarily
$c_1 < \dots <c_K$.
\end{lemma}


\noindent{\textbf{Proof of Lemma \ref{lem:differentCentroidsOptim}}:}
 Ad absurdum, we suppose that  an optimal solution exists with centroids such that $c_{k'}= c_k$.
Having  $K < N$, there exist a point $x_n$ that is not a centroid (note that points of $E$ are distinct in the hypotheses of this paper).
Merging clusters $\CC_{k'}$ and $\CC_k$ does not change the objective function as the centroid  are the same. Removing $x_n$ from its cluster and defining it in a singleton cluster strictly decreases the objective function,
it is a strictly better solution than the optimal solution. \hfill$\square$

\vskip 0.2cm

\noindent{\textbf{Proof of Proposition \ref{prop:intervalClusteringPMSSC}}:}
Let $X$ the set of selected outliers in an optimal solution of weighted PMSSC.
Ad absurdum, if there exists a strictly better solution of weighted MSSC in $E \setminus X$, adding $X$ as outliers would imply a strictly better solution for PMSSC in $E$, in contradiction with the global optimality of the given optimal solution. Lemma \ref{lem:intervalClusteringMSSC} holds in $E \setminus X$. \hfill$\square$

%

\vskip 0.2cm

\noindent{\textbf{Proof of Proposition \ref{prop:intervalClusteringUnweightedPMSSC}}:}
Let $X$ the set of selected outliers in an optimal solution of unweighted PMSSC.
Ad absurdum, we suppose that there exists a cluster $\CC$ of centroid $c$ with $x_j = \min \CC$,
$x_{j'} = \max \CC$ and $x_i \in X$ such that $x_{j} < x_{i} < x_{j'}$.
If $c \leqslant x_{i}$, the objective function strictly decreases when swapping $x_{i}$ and $x_{j'}$
in the cluster and outlier sets.
If $c \geqslant x_{i}$, the objective function strictly decreases when swapping $x_{i}$ and $x_{j}$
in the cluster and outlier sets. This is in contradiction with the global optimality.
For the end of the proof, let us count the outliers. We have:
$i_1-1 + i_2-1 -j_1 + ... + i_{K}-1 -j_{K-1} + N - j_{K} \leqslant M $ which is equivalent to
 $ N + \sum_{k=1}^{K}  (i_k-j_k)  \leqslant M + K$. \hfill$\square$

\vskip 0.2cm
\noindent{\textbf{Proof of Proposition \ref{prop:inductionCostCluster}}}: Relations
 (\ref{inducCumulatedWeight1}) and (\ref{inducCumulatedWeight2}) are trivial with the definition of $v_{i,i'}$ as a sum.
 Relations
 (\ref{inducCentroid1W}) and (\ref{inducCentroid2W}) are standard associativity relations with weighted centroids.
We prove here  (\ref{inducCost2W}), the proof of (\ref{inducCost1W}) is similar.
%
%
%
\footnotesize
$c_{i-1,i'} - w_{i-1}(x_{i-1} - b_{i-1,i'})^2 = \sum_{j=i}^{i'} w_j (x_j - b_{i-1,i'})^2$\\
 $=  \sum_{j=i}^{i'} w_j \left ( (x_j - b_{i,i'})^2 + (b_{i,i'} - b_{i-1,i'})^2 + 2  (x_j - b_{i,i'}) (b_{i,i'} - b_{i-1,i'}) \right)$\\
$= c_{i,i'} +  (b_{i,i'} - b_{i-1,i'})^2 \sum_{j=i}^{i'} w_j +2   (b_{i,i'} - b_{i-1,i'})  \sum_{j=i}^{i'} w_j (x_j - b_{i,i'})$.\\
 \normalsize
 It gives the result as $ \sum_{j=i}^{i'}  w_j (x_j - b_{i,i'}) =   \sum_{j=i}^{i'}  w_j x_j - b_{i,i'} \sum_{j=i}^{i'} w_j  =0$. \hfill$\square$

\vskip 0.2cm

\noindent{\textbf{Proof of Proposition \ref{prop:computationCost1}}:}
We compute and store values $c_{1,i}$  with $i$ increasing starting from $i=1$.
We initialize $v_{1,1} = w_1$, $b_{1,1} = x_1$ and $c_{1,1}=0$ and compute values
$c_{1,i+1},b_{1,i+1},v_{1,i+1} $ from $c_{1,i},b_{1,i},v_{1,i}$ using (\ref{inducCost1W})  (\ref{inducCentroid1W})
 (\ref{inducCumulatedWeight1}). Such computation is in $O(1)$ time, so that
  cluster costs $c_{1,i}$ for all $i \in [\![1;N]\!]$ are computed in $O(i)$ time.
  In memory, only four additional elements are required: $b_{1,i+1},v_{1,i+1},b_{1,i},v_{1,i}$. The space complexity is given by the stored
  $c_{1,i}$ values.  \hfill$\square$

\vskip 0.2cm
\noindent{\textbf{Proof of Proposition \ref{prop:computationCost2}}:} Let   $j \in [\![1;N]\!]$.
We compute and store values  $c_{i,j}$  with $i$ decreasing starting from $i=j$.
We initialize $v_{j,j} = w_j$, $b_{j,j} = x_j$ and $c_{j,j}=0$ and compute values
$c_{i-1,j},b_{i-1,j},v_{i-1,j} $ from $c_{i,j},b_{i,j},v_{i,j}$ using (\ref{inducCost2W})  (\ref{inducCentroid2W})
 (\ref{inducCumulatedWeight2}). Such computation is in $O(1)$ time, so that
  cluster costs $c_{i,j}$ for all $i \in [\![1;j]\!]$ are computed in $O(j)$ time.
  In memory, only only four additional elements are required $b_{i-1,j},v_{i-1,j}, b_{i,j},v_{i,j}$ the space complexity is in $O(i)$.  \hfill$\square$

\vskip 0.2cm
  \noindent{\textbf{Proof of Proposition \ref{prop:complexityUnweightedK=1}}:}
Using interval optimality for unweighted PMSSC, we compute successively $c_{1,N-M}, c_{2,N-M+1}, \dots, c_{M+1,N}$ and store the best solution.
Computing $c_{1,N-M},b_{1,N-M},v_{1,N-M} $ is in $O(N-M)$ time with a naive computation.
Then  $c_{1,N-M+1},b_{1,N-M+1},v_{1,N-M+1} $ are computed from  $c_{1,N-M},b_{1,N-M},v_{1,N-M} $
in $O(1)$ time using successively (\ref{inducCumulatedWeight1}),
(\ref{inducCentroid1W}) and (\ref{inducCost1W}).
Then  $c_{2,N-M+1},b_{2,N-M+1},v_{2,N-M+1} $ are computed  from $c_{1,N-M+1},b_{1,N-M+1},v_{1,N-M+1} $ in $O(1)$ time using successively (\ref{inducCumulatedWeight2}),
(\ref{inducCentroid2W}) and (\ref{inducCost2W}).
This process is repeated $M$ times, there are $O(N-M)$ + $O(M)$ operations, it runs in $O(N)$ time.
Spatial complexity is in $O(1)$.  \hfill$\square$

\vskip 0.2cm
\noindent{\textbf{Proof of Proposition \ref{prop:complexityWeightedK=1M=1}}:}
We enumerate the different costs considering all the possible outliers.
We compute  $c_{1,N},b_{1,N},v_{1,N} $ in $O(N)$ time.
Adapting Proposition \ref{prop:intervalClusteringUnweightedPMSSC}, each cluster cost removing
one point can be computed in $O(1)$ time. The overall time complexity is in $O(N)$.
 \hfill$\square$

\vskip 0.2cm
\noindent{\textbf{Proof of Proposition \ref{bellman}}}:
 (\ref{inducInit0}) is the standard  case $K=1$.
  (\ref{inducInit2}) is a trivial case where  the optimal clusters are singletons. 
   (\ref{inducInit3}) is a recursion formula among  $K=1$ cases, either point $x_i$ is chosen and in this case all the outliers are points
   $x_l$ with $l  \leqslant m$ or $x_i$ is an outlier and it remains an optimal PMSSC with $K=1$ and $M=m-1$ among the $i-1$ first points.
    (\ref{inducForm}) is a recursion formula among   $M=0$ cases distinguishing the cases for the composition of the last cluster.
 (\ref{inducForm})  are considered for MSSC in \cite{gronlund2017fast,wang2011ckmeans}.
(\ref{inducForm2}) is an extension of  (\ref{inducForm}). $O_{i,k,m}$ is $O_{i-1,k,m-1}$ if point $x_i$ is not selected.
Otherwise, the cluster $k$ is a $\CC_{j,i}$ and the optimal cost of other clusters is $O_{j-1,k-1,m}$.   \hfill$\square$

\vskip 0.2cm
\noindent{\textbf{Proof of Theorem \ref{Thm1}}:  by induction, one proves that at each loop $i$ of In Algorithm 1, the optimal values of  $O_{i,k',m'}$ are computed for all $k',m'$ using Proposition \ref{bellman}.  Space complexity is given by the size of DP matrix $(O_{i,k,m})$, it is in $O(KN(1+M))$.
 Each value requires at most $N$ elementary operations,  building  the DP matrix runs in $O(KN^2(1+M))$ time. The remaining of Algorithm 1 is a standard backtracking procedure for DP algorithms, running in $O(N^2)$ time, the time complexity of DP is thus in $O(KN^2(1+M))$.
 Lastly, unweighted PMSSC is polynomially solvable with Algorithm 1, the memory space of inputs are in $O(N)$, mostly given y the $N$ points of $E$, and using $K,M \leqslant N$, the time and  space complexity are respectively  bounded by  $O(N^4)$ and $O(N^3)$. \hfill $\square$}

\end{document}